\documentclass[fleqn,usenatbib]{mnras}

\usepackage{newtxtext,newtxmath}
\usepackage{hyperref}
\usepackage{xcolor}
\usepackage{subcaption}
\usepackage[T1]{fontenc}
\usepackage{ae,aecompl}
\usepackage{etoolbox}
\makeatletter
\patchcmd\@combinedblfloats{\box\@outputbox}{\unvbox\@outputbox}{}{%
   \errmessage{\noexpand\@combinedblfloats could not be patched}%
}%
 \makeatother

\usepackage{graphicx}	
\usepackage{amsmath}	
\usepackage{amssymb}	

\title[Morphological classification of galaxies in CLASH: using deep neural networks]{Multiband galaxy morphologies for CLASH: a convolutional neural network transferred from CANDELS}

\author[M. P\'erez-Carrasco et al.]{
M. P\'erez-Carrasco,$^{1}$\thanks{E-mail: maperezc@udec.cl}
G. Cabrera-Vives, $^{1,2}$ 
M. Martinez-Marin,$^{3}$
P. Cerulo,$^{3}$
\newauthor{ R. Demarco,$^{3}$
P. Protopapas,$^{4}$
J. Godoy,$^{1}$
and M. Huertas-Company $^{5,6}$}
\\
$^{1}$Department of Computer Science, University of Concepci\'on, Casilla 160-C, Concepci\'on, Chile\\
$^{2}$Millennium Institute of Astrophysics, Santiago, Chile\\
$^{3}$Department of Astronomy, Universidad de Concepci\'{o}n, Casilla 160-C, Concepci\'{o}n, Chile\\
$^{4}$Institute for Applied Computational Science, Harvard University Northwest B162, 52 Oxford Street Cambridge, MA 02138, USA \\
$^{5}$LERMA, Observatoire de Paris, PSL Research University, CNRS, Sorbonne Universit\'es, UPMC Univ. Paris 06, F-75014 Paris, France \\
$^{6}$University of Paris Denis Diderot, University of Paris Sorbonne Cit\'e (PSC), 75205 Paris Cedex 13, France
}

\date{Accepted XXX. Received YYY; in original form ZZZ}

\pubyear{2015}

\begin{document}
\label{firstpage}
\pagerange{\pageref{firstpage}--\pageref{lastpage}}
\maketitle

\begin{abstract}
We present visual-like morphologies over 16 photometric bands, from ultra-violet to near infrared, for $8,412$ galaxies  in the Cluster Lensing And Supernova survey with Hubble (\textit{CLASH}) obtained by a convolutional neural network (CNN) model. Our model follows the CANDELS main morphological classification scheme, obtaining the probability for each galaxy at each CLASH band of being spheroid, disk, irregular, point source, or unclassifiable. Our catalog contains morphologies for each galaxy with $H_{mag}<24.5$ in every filter where the galaxy is observed. We trained an initial \textbf{CNN} model using approximately 7,500 expert eyeball labels from The Cosmic Assembly Near-IR Deep Extragalactic Legacy Survey (CANDELS). We created eyeball labels for 100 randomly selected galaxies per each of the 16-filters set of CLASH (1,600 galaxy images in total), where each image was classified by at least five of us. We use these labels to fine-tune the network in order to accurately predict labels for the CLASH data and to evaluate the performance of our model.
We achieve a root-mean-square error of $0.0991$ on the test set. We show that our proposed fine-tuning technique reduces the number of labeled images needed for training, as compared to directly training over the CLASH data, and achieves a better performance. This approach is very useful to minimize eyeball labeling efforts when classifying unlabeled data from new surveys. This will become particularly useful for massive datasets such as the ones coming from near future surveys such as EUCLID or the LSST. Our catalog consists of prediction of probabilities for each galaxy by morphology in their different bands and is made  publicly available at \url{http://www.inf.udec.cl/~guille/data/Deep-CLASH.csv}.

\end{abstract}

\begin{keywords}
Galaxies -- Morphology -- Classification -- Catalog
\end{keywords}



\section{Introduction}
Galaxies are complex systems and understanding their evolution represents one of the key questions in modern astrophysics. The physical properties of a galaxy, such as its baryonic content (stellar, gas and dust masses), star formation rate (SFR), structure (morphology) and chemical abundance, change over time and their evolution is driven by a combination of internal and environmental processes (\citealt{Kauffmann_2003,Kauffmann_2004}; \citealt{Peng_2010}; \citealt{Muzzin012}). The morphology of galaxies, which is related to their structural and dynamical properties and to their star formation history (quantified by their SFR over time), represents a fundamental and powerful diagnostic to study evolutionary changes of galaxies.

Since the first attempts to understand these `nebulae', astronomers have been classifying galaxies according to their visual appearance. In 1926, E.\ P. Hubble proposed a tuning-fork scheme \citep{Hubble} that would constitute the basis for any morphological classification until present. In the tuning-fork diagram, galaxies are separated into early and late morphological types, the former including galaxies without disks and spiral arms and the latter comprising all galaxies that showed spiral arms, with a spheroidal or elliptical light distribution, and in their disks. In Hubble's scheme there is also a class of lenticular galaxies, also known as S0, which correspond to systems made up of a central bulge structure surrounded by a disk component but without spiral arms. Finally, a class of irregular galaxies, which do not show any prominent morphological signature (e.g.\ the Magellanic clouds), have also been defined and considered as a late morphological type.

This bi-modality in galaxy morphology provides us information about the structural distribution of their stellar composition and therefore the processes that shaped them, including stellar mass assembly, galaxy-galaxy interactions and other environmental effects. One of the open questions is whether there is any clear evolutionary link between early and late-type galaxies. 

To address this question, astronomers have classified galaxies in samples of varying sizes and built morphological catalogs. Traditionally, galaxies have been classified by visually inspecting high-resolution images of varying depths and sizes (e.g.\ \citealt{De_Vaucouleurs_1991}, \citealt{Dressler1,Dressler2}, \citealt{Couch_1998}, \citealt{postman}, \citealt{Desai_2007}, \citealt{Nair_2010}). Usually, more than one classifier is employed in this task. Although the human eye is able to distinguish a great amount of details, such as spiral arms, bulges and bars, visual classification has gradually become impractical as the volume of data delivered by astronomical surveys has grown significantly. Astronomers have therefore passed from facing themselves the classification of a few hundreds of galaxies to classify tens to hundreds of thousands of galaxies (\citealt{Nair_2010}, \citealt{karteltepe}). 

In the last decade, the morphological classification of galaxies has been addressed from two complementary points of view, namely, visual classification by a large number of non-expert people through crowd-sourcing platforms (e.g.\ Galaxy Zoo and Galaxy Zoo CANDELS, \citealt{Lntott_2008}, \citealt{Simmons_2017}) and automated classification through machine learning algorithms. The latter tries to find a set of physical and observational parameters that correlate with the visual morphology of a galaxy and defines the space of parameters that best characterizes a given morphological type (\citealt{abraham}, \citealt{conselice}, \citealt{lotz}, \citealt{Huertas_2008, Huertas_2011}). These methods suffer from difficulties in reaching the levels of reliability required for scientific analysis and their uncertainties remain high when one tries to go beyond the early- vs late-type scheme and to distinguish elliptical from lenticular galaxies (see e.g.\ discussion in \citealt{Cerulo_2017}). 

In recent years, deep learning methods that mimic the human eye perception were able to learn the best set of parameters for a given problem. \cite{Dieleman} (hereafter D15) trained convolutional neural networks (\textit{ConvNets}; \citealt{Fukushima}, \citealt{Lecun}), a deep learning model, retrieving Galaxy Zoo's visual classifications from Sloan Digital Sky Survey (SDSS, \citealt{York_2000_SDSS}) images with $>99$\% accuracy. A similar method was later applied at higher redshifts in the Cosmic Assembly Near-infrared Deep Extragalactic Legacy Survey (CANDELS, \citealt{grogin011}) by \cite{marc} (see \S 2 for details).

Motivated by these works we decided to apply a ConvNet-based classification to galaxies in the Cluster Lensing and Supernova Survey with Hubble (CLASH, \citealt{Postman_2012}), a survey of 25 clusters of galaxies at redshifts $0.2 < z < 0.8$ observed in up to 16 photometric bands with the Advanced Camera for Surveys (ACS) and the Wide Field Camera 3 (WFC) on board the Hubble Space Telescope (HST). CLASH provides us with publicly available deep and high-resolution (0.065$''/$pixel or 0.03$''/$pixel) images of massive cluster cores with a wavelength coverage that spans the electromagnetic spectrum from the near Ultraviolet (NUV, 0.2 $\mu m$) all the way through to the near Infrared (NIR, $1.6 \mu m$). CLASH constitutes the optimal counterpart to deep field surveys such as CANDELS for environmental studies of galaxies at intermediate redshifts, because clusters of galaxies are the most massive, virialized large-scale structures in the Universe and host a broad variety of environments, from the dense cores to the sparser and dynamically active outskirts. All this makes them observational laboratories for the study of the environmental processes that drive the evolution of galaxies. Galaxy clusters are characterized by a morphology-density relation (\citealt{Dressler1}) up to $z \approx 1.5$. More precisely, it is observed that early-type galaxies in clusters are more frequent in their inner regions, while the fraction of late-type galaxies increases towards the cluster outskirts (\citealt{Dressler2}, \citealt{postman}, \citealt{hilton09}, \citealt{Muzzin012}, \citealt{mei012}, \citealt{Holden_2007}). It is also observed that the fraction of blue, star-forming galaxies in clusters increases with redshift (\cite{bo1978}), suggesting that clusters promote the suppression of star formation and the formation of early-type galaxies with time (see also \citealt{Fasano_2001}). 

\begin{figure}
  \centering
\includegraphics[width=8.5cm]{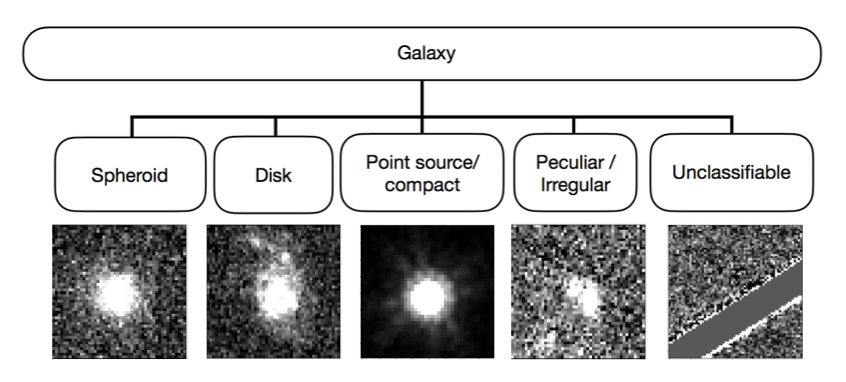}
  \caption{Morphology classification scheme used to classify galaxies in CLASH, from KA15. Images show examples for each class in the CLASH f160w filter.}
  \label{scheme}
\end{figure}

To classify galaxies in CLASH we train a convolutional neural network (ConvNet) architecture based on Inception \citep{Szegedy} using CANDELS F160W (H band, $1.6 \mu m$) images of galaxies in the GOODS-S field that were visually classified by \cite{karteltepe} (KA15 hereafter), obtaining a root-mean square error (RMSE) of $\sim0.125$. Then, we fine-tuned our model to the CLASH data using the previous knowledge acquired from CANDELS. We predict labels on CLASH galaxies in each of the 16 available HST photometric bands independently (F225W, F275W, F336W, F390W, F435W, F475W, F606W, F625W, F775W, F814W, F850LP, F105W, F110W, F125W, F140W, F160W) obtaining a $0.0991$  RMSE over a subsample of CLASH galaxies visually labeled according to the classification scheme in Figure \ref{scheme}.
We apply our model to $8,412$ galaxies from CLASH and release the first multi-band morphological catalog of CLASH galaxies.

This paper is organized as follows: in \S 2 we provide a description of the principles and functioning of ConvNets and Transfer Learning; in \S 3 we discuss our training sample and labeling scheme; in \S 4 we present our methodology; in \S 5 we present our results and discussion; in \S 6 we present the catalog; finally in \S 7 we summarize and conclude our work.

\section{TECHNICAL BACKGROUND AND PREPARATORY WORK}

\subsection{Convolutional Neural Networks}

A convolutional neural network (\textit{ConvNet}) (\citealt{Fukushima}, \citealt{Lecun}) is a deep learning model aimed at modeling the data, through different abstraction levels, using convolutional stacks followed by non linearities and statistical aggregation operations (e.g.\ maxpooling; \citealt{maxpool}). The resulting features are connected to a prediction model, such as a feed-forward neural network \citep[see][and references therein]{zhang2000}, to obtain responses for a learning task. They are commonly used when the data exhibit a given kind of topological structure that needs to be preserved. This is the case of the pixels in an image. 

The parameters of the model are iteratively learnt by minimizing a loss function for subsets of the data, called mini-batches. The parameters of the model are updated for each mini-batch of training examples using backpropagation \citep{Lecun}. Backpropagation is an efficient algorithm to compute the partial derivatives of the loss function with respect to the parameters of each layer of the model, in order to iteratively modify such parameters using a gradient descent method. 
Stochastic gradient descent (SGD) updates the parameters in the opposite direction of the gradients modulated by a \textit{learning rate}.
Adam \citep{ADAM_2014} is a stochastic optimization algorithm used to update weights in which the learning rate changes during training. The way in which Adam computes learning rates is through an exponentially decaying average of past gradients and the squared decaying averages of past gradients \citep{ruder2016}, 
and then control the decay rates of these moving averages using a bias correction.

Despite ConvNets been introduced a long time ago, the absence of computational power, enough amount of data and efficient regularization algorithms did not allow the successful implementation of ConvNet architectures until recently. Only in 2012, thanks to the use of ReLU non linearities (\citealt{Nair}) and Dropout regularization (\citealt{hinton}, \citealt{dropout}), the architecture proposed by \cite{Krizhevsky} obtained a considerably advantage in the ImageNet Classification Challenge, one of the most important contests in the field of computer vision, significantly improving over the previous state-of-the-art algorithms.

Since then, the most successful image classification algorithms have been based on ConvNets (\citealt{simon}, \citealt{Szegedy}, \citealt{He}), using deeper architectures and methods which allow better backpropagation of the errors (e.g.\ batch normalization, see \citealt{batchnorm} for details).

In the context of astronomical images, the first attempts to generate classification algorithms using ConvNets were made by \cite{Dieleman} for Galaxy Zoo - The Galaxy Challenge (posted in Kaggle \footnote{\url{https://www.kaggle.com/c/galaxy-zoo-the-galaxy-challenge}}), which tried to find the best model for morphological classification of galaxies in SDSS/Galaxy Zoo\footnote{Galaxy Zoo is a crowd-sourcing in which expert and non expert people could classify jpg galaxies by their morphology.} labeled images. They obtained an accuracy higher than $99$\%. 
The second attempt was made by \cite{marc} who trained a model to morphologically classify galaxies in 5 fields of the CANDELS survey. They obtained a misclassification error lower than $1$\%.

More recently, deep learning was used by \cite{aniyan_2017} to classify radiogalaxies on images from the Very Large Array (VLA), while \cite{Dominguez_2018} ran a deep learning algorithm to produce a detailed morphological classification of 670,000 SDSS galaxies. Other fields of application of deep learning in astronomy include detection of extrasolar planets, the study of transients, the characterization of pulsars and fast radio bursts (FRB) (\citealt{Shallue_2018}, \citealt{Guille_2017}, \citealt{Guo_2017}, \citealt{connor_2018}). Deep learning have also been used for the estimation of photometric redshifts in astronomical surveys with limited photometric information (e.g. KIDS; \citealt{petrillo_2017}).

\subsection{Fine-tuning ConvNet architectures:}
In real world applications, collecting enough amount of labels to train deep learning models is expensive and time-consuming. In some cases, 
a set of labeled images from a related domain (\textit{source}) could be used as an initial learning set, whose obtained parameters could then be used to train a model with few labeled data (\textit{target}). This is called \textit{transfer learning}  or \textit{domain adaptation} \citep{pan_2010}.


Deep learning models are characterized by learning mid-level representations of the images in each convolutional layer (\citealt{yosinskiunder}, \citealt{zeiler}) that show general abstractions in the first layers, while becoming more specific in the last layers. This allows the use of the first layers directly from the source and to train only layers which map more specific details of the images. This is called \textit{fine-tuning} (\citealt{yosinskifine}; \citealt{oguab}).

The general suggested  fine-tuning approach, when related domains are available (i.e., galaxies from different surveys), is to copy layers trained on source, keeping frozen some of these layers and train from last layer frozen ahead on the target (\citealt{Chu_2016}, \citealt{yosinskifine}). The layers to be frozen will depend on the problem and the constructed architecture. In this paper, we evaluate the performance of our model in terms of the number of layers frozen.

\section{DATA}

\subsection{Images}

We trained a baseline model on CANDELS images and transfered it to CLASH images. To train our baseline model, we used HST images from a CANDELS field taken with WFC3 in the F160W band, namely GOODS-S \citep{giava04}\footnote{The Great Observatories Origins Deeps Survey}. Notice that we are not using the following CANDELS fields: COSMOS \citep{scov07} , UDS \citep{law07} , and EGS \citep{davis07}, as the labels of KA15 come from GOOD-S. To these we added the mosaic from Hubble Legacy Fields (HLF) Data Release 1.5 for the GOODS-S region (HLF-GOODS-S) (Illingworth, Magee, Bouwens, Oesch et al, 2017, in preparation), which combines exposures from Hubble's  Advanced Camera for Surveys Wide Field Channel (ACS/WFC) and the Wide Field Camera 3 InfraRed Channel (WFC3/IR). 

For both sets of images we used the 0.06$''/$pixel resolution version in the filter F160W, selecting galaxies with F160W magnitudes $H_{mag}<24.5$. This is the same limit used in KA15, who show that it corresponds to the flux limit for reliable visual morphological classifications.
We created postage-stamp images from the GOODS-S mosaic setting the size to four times the Petrosian radius as reported in the catalog of \cite{guo013}. 
We then perform a bi-lineal interpolation to set all images sizes to 80$\times$80 pixels. 
We obtain a final sample of $\sim7,500$  galaxies.

For the transfer learning sample we used images from the CLASH Multi-Cycle Treasury program (\citealt{postman}). CLASH observed 25 clusters of galaxies at redshifts $0.15 < z < 0.9$ with WFC3 over a period of 3 years, in up to 16 filters, namely F225W, F275W, F336W, F390W, F435W, F475W, F606W, F625W, F775W, F814W, F850W, F105W, F110W, F125W F140W and F160W, covering the ultra violet (UV), optical (OPT) and NIR regions of the spectrum. \cite{molino017} published accurate multiwavelength photometric catalogs for these clusters which also provide the Petrosian radius. We created postage-stamp images for each filter separately following the same criterion for the magnitude cut and the size that we adopted for CANDELS ending up with a sample of $68,531$ galaxies.

\subsection{Labels}

In order to fine-tune and evaluate our model over CLASH data, we created a limited set of labels for CLASH images. We used a scheme similar to \cite{karteltepe} for the classification of galaxies in CANDELS, which considers spheroids, disks, point sources, irregulars, and unclassifiable sources (see Figure \ref{scheme}). 
We randomly selected 100 galaxies in each CLASH band (1,600 in total) and classified them by eye. Each galaxy was labeled in each band by at least five of us on gray-scale images with  $95\%$ and $99.5\%$ stretches that were uploaded on the Zooniverse web platform\footnote{\url{https://www.zooniverse.org}}. Notice each human annotator produced up to 16 labels per galaxy, not necessarily the same across bands, as our goal is to have a model that produces different classifications for each band. At the end of this process a probability for the galaxy to have a certain morphological type was assigned to each object. This probability is defined as:
\begin{equation}
P_{T} = \frac{N_T}{N_{\mathrm{tot}}}, 
\end{equation}
where $N_T$ is the number of people who assigned a type $T$ to the galaxy and $N_{\mathrm{tot}}$ is the total number of people who classified that galaxy. We will call this labeled dataset CL-eye hereafter.

\section{DEEP LEARNING FRAMEWORK}

Our model predicts the probability of each galaxy to be of each type, and it is based on the Inception model designed by \citealt{Szegedy}. We implemented it on the Keras deep learning library \citep{keras} with a TensorFlow backend \citep{tensorflow}. The more galaxies we use for training our ConvNet model the best performance it will achieve \citep{Chu_2016}. Our goal is to have an accurate predicting model for morphology probabilities over CLASH single-filter images. However, since before the making of this catalog no labels were available for morphologies of galaxies in CLASH, we used the KA15 CANDELS morphology catalog and trained our model with this dataset. We later fine-tuned the parameters of the model to a smaller subset of galaxies that we labeled for CLASH.


\subsection{Architecture}

\begin{figure*}
\begin{subfigure}[t]{0.32\textwidth}
\center
\includegraphics[width=1\textwidth]{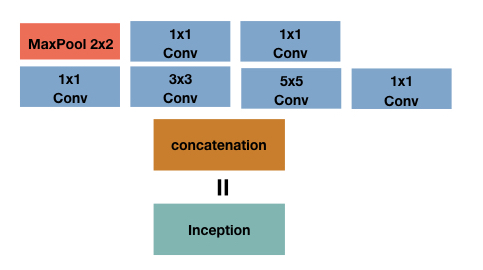}
\caption{Inception module}
\label{incepmod}
\end{subfigure}%
\hfill
\begin{subfigure}[t]{0.67\textwidth}
\includegraphics[width=1\textwidth]{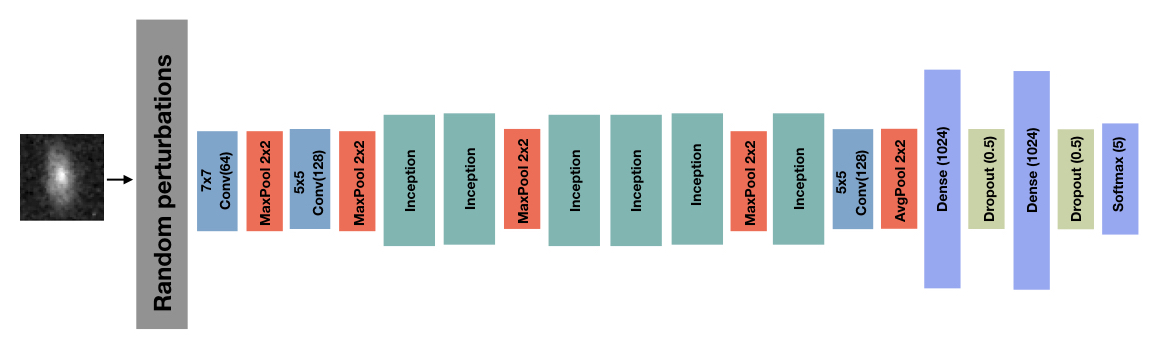}
\caption{ConvNet architecture}
\label{arch}
\end{subfigure}
        \caption{ a) Example of an inception module which is composed by 3 convolutions with different sizes and a maxpooling operation, b) Convolutional Neural Network architecture based on Inception (\protect\citealt{Szegedy}) used in this work to train a model over the $\sim7,500$ galaxies from GOOD-S field of CANDELS catalog by KA15 and fine-tuned over CLASH data. The training starts with images of size 80x80 pixels which are randomly perturbed meanwhile training and convolving over six inception modules as showed, getting a reduced set of features in the final stack. After that, the low level features are passed through a Neural Network with two hidden layers in order to predict the value for every possible type of galaxy using a softmax activation function.}
        \label{Closing}
\end{figure*}

In general, when a ConvNet architecture is used to predict image features, one of the most important hyperparameters are the filter sizes used to perform the convolutions. The inception model proposed by \citealt{Szegedy} combines different convolutional filters into a single unit. It first makes a $1\times1$ convolution reducing the dimensionality of the feature space and then uses different filter sizes in the same layer paddings to maintain dimensions. Resulting operations are concatenated at the end of the inception module (see Figure \ref{incepmod}).
 
The model that we used is composed of six inception modules that map images of galaxies into a set of $1,024$ features grouped in the final convolutional stack. We use batch-normalization \citep{batchnorm} in all layers before the fourth inception unit. These features are passed through three dense layers of $1,024$, $1,024$, and 5 neurons that represent every possible value that the model can take for every type of galaxy using a softmax activation function (see Figure \ref{arch}). Notice that our model has $\sim 5.5$ million of parameters to be fitted.  Details about the hyperparameters used in this architecture are shown in Table \ref{hyper}. 

\begin{table*}
    \begin{center}
    \begin{tabular}{ | c | c | c | c | c | c | c | c | c |}
    \hline
   Operation  & Filter size / & Output Size  & $1\times1$  & $3\times3$ & $1\times1$ & $5\times5$ & $1\times1$ & $1\times1$\\  & depth &   & before $3\times3$ & &  before $5\times5$ &  &  & after maxpooling\\  \hline 
     convolution& $7\times7$ / $64$ & $74\times74\times64$ & & & & & & \\ \hline
     max pool & &$37\times37\times64$ & & & & & &  \\ \hline
     convolution& $6\times6$ / $128$ &$32\times32\times128$ & & & & & &   \\ \hline
     max pool & &$16\times16\times128$ & & & & & \\ \hline
     inception & &$16\times16\times256$ &$96$&$128$&$16$&$16$&$64$&$32$ \\ \hline
     inception & &$16\times16\times480$&$128$&$192$&$32$&$96$&$128$&$64$\\ \hline
     max pool & &$8\times8\times480$ & & & & & \\ \hline
     inception & &$8\times8\times512$&$96$&$208$&$16$&$48$&$192$&$64$ \\ \hline
     inception & &$8\times8\times512$&$112$&$224$&$24$&$64$&$128$&$64$ \\ \hline
     inception & &$8\times8\times512$&$144$&$288$&$32$&$64$& $128$&$64$\\ \hline
     max pool & &$4\times4\times512$ & & & & & \\ \hline
     inception & &$4\times4\times528$&$144$&$288$&$32$&$64$& $112$&$64$\\ \hline
     avg pool & $2\times2$ &$2\times2\times528$ & & & & & \\ \hline
     convolution & $1\times1$ / $256$ &$2\times2\times256$ & & & & & \\ \hline
     flatten &  &$1\times1\times1024$ & & & & & \\ \hline

     dense & &$1\times1\times1024$ & & & & & \\ \hline
     dropout (0.5) & &$1\times1\times5$ & & & & & \\ \hline

     dense & &$1\times1\times1024$ & & & & & \\ \hline
     dropout (0.5) & &$1\times1\times5$ & & & & & \\ \hline

     softmax & &$1\times1\times5$ & & & & & \\ \hline
    \end{tabular}
    \caption{Architecture of our model. The table follows the same order of Fig. \ref{arch}. }
     \label{hyper}
    \end{center}
\end{table*}

\subsection{Training strategy}\label{sec:training}

Taking advantage of the rotation-invariant property of the galaxies (see D15), during training time, in each epoch we artificially augmented the number of images in the training set four times by applying random perturbations:

\begin{itemize}
\item \textbf{rotations}: random rotations are performed by sampling the rotation angle from a uniform probability distribution with values between 0 and 360.
\item \textbf{flipping}: the images are flipped horizontally and vertically with a probability of 0.5 each.
\item \textbf{translation}: translations are performed by sampling two values from a uniform probability distribution on the interval [-4, 4], representing the number of pixels to be translated in the $x$ and $y$ coordinates of the image.
\end{itemize} 
With these perturbations it is unlikely that our model sees exactly the same image twice while training. This data augmentation helps avoiding overfitting. 

We perform validation by dividing the sample in 3 subsets without overlapping: a training set, a validation set, and a test set. We use the training set to backpropagate the gradients and evaluate using the validation set (without random perturbations) at training time in order to test for convergence. Finally, we report the test error measured by evaluating the fully trained model over the test set (without random perturbations). This way, we have a final evaluation of our model over data never used at training time.

In order to compute the weights of the architecture used to predict the labels, we use the ADAM method for stochastic optimization \citep{ADAM_2014}. In order to compute the weights of the architecture used to predict the labels, we use the ADAM method for stochastic optimization \citep{ADAM_2014}. We tried different values for the learning rate and obtained the best results in terms of cross validation RMSE by using a learning rate of $5\times10^{-5}$ \footnote{We evaluated for learning rates of [$1\times10^{-5}$, $5\times10^{-5}$, $1\times10^{-4}$, $1\times10^{-4}$, $1\times10^{-3}$].}. We used initial values for $\beta_{1}$ and $\beta_{2}$ of 0.9 and 0.999 respectively. In addition, the model was also trained using cross-entropy as loss function, achieving statistically similar results to using RMSE in terms of test evaluation.

\section{Performances and Discussion}
We train the model using $\sim7,500$ galaxies from the CANDELS GOODS-S catalog of KA15 in order to acquire prior knowledge to train the model with CLASH data. Our work was divided in two steps, namely prior knowledge acquisition from CANDELS and fine-tuning over CLASH. In this section, we explain our methodology in detail.

\subsection{Prior knowledge acquisition from CANDELS}

 We trained the ConvNet model described in Fig. \ref{arch} using CANDELS data and labels from KA15. We split the data into a 70\% for training, a 15\% for validation, and a 15\% for testing the final model. We used 160 mini-batches of 128 images per epoch, in which each image was randomly sampled with replacement, and we applied to it the perturbations described in \S \ref{sec:training}. The evaluation of the model, was made in real-time using images from the validation set after each epoch. We backpropagate the gradients on our training set as long as the validation test loss diminishes within 35 epochs. 

The learning curve of the model trained over CANDELS is shown in Figure \ref{loss}. We obtain a RMSE of $\sim0.123$ on the test set, consistent with results from \cite{marc}. This model is used as prior knowledge to train a model over CLASH data.

\subsection{Fine-tuning over CLASH}

 After training the model over the CANDELS data using the labels from KA15, we fine-tuned it to the CLASH data using the labels that we generated. In order to assess the need of fine-tunning, we also evaluated our model trained directly on CLASH labels.

\subsubsection{Training CLASH without fine-tuning}

 We started by training our model directly on CL-eye data. 
We used a validation set of 150 galaxies and a test set of 150 galaxies. Some galaxies in CL-eye were labeled more than once on different filters. The same galaxy was never present in more than one of these subsets. Each training epoch contained 30 mini-batches of 128 images per epoch. Recall each image in the training mini-batches was randomly sampled with replacement, and random perturbations were applied to them. 

\begin{figure}
  \centering
\includegraphics[width=8cm]{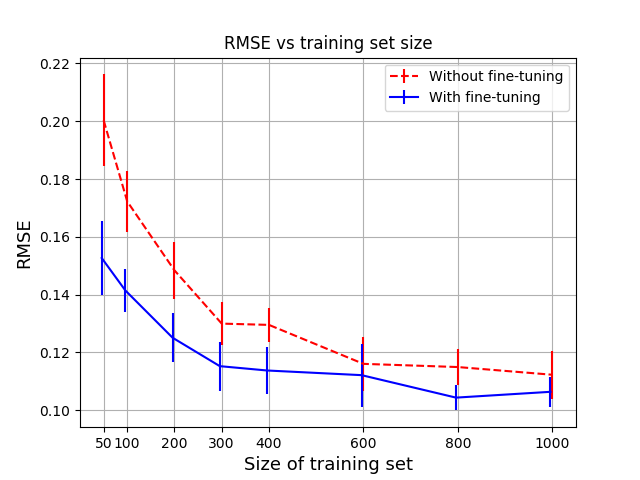}
  \caption{RMSE loss in terms of the number of galaxies in the training set. Error bars show the RMSE standard deviation for 10 random train-validation-test cross-validation splits.}
  \label{lossvssize}
\end{figure}

In order to assess the amount of labeled data needed to train our models, we trained different models using training sets of 50, 100, 200, 300, 400, 600, 800 and 1000 images. The results of 10 cross validation runs for each training size are shown in Figure \ref{lossvssize}. As expected, as the training size increases, the test set loss goes down. When using more than 600 objects for training, there is no statistically significant improvement. 

\begin{figure}
  \centering
\includegraphics[width=8cm]{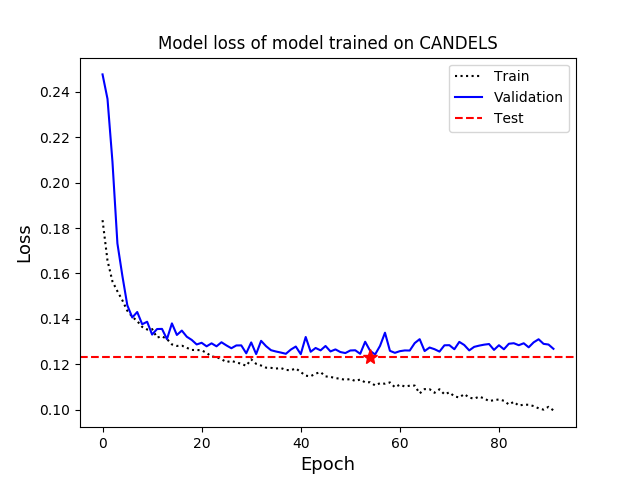}
  \caption{Learning curve of the model trained and evaluated using CANDELS data. The black dotted line shows the loss over the training set, the blue solid line shows the loss over the validation set and the dashed red line shows the final test set loss over the best validation set model. A red star is shown at the epoch such model was obtained. The loss function shown is the RMSE between KA15 labels and predictions given by the model.}
  \label{loss}
\end{figure}

\subsubsection{Training CLASH with fine-tuning}

 Here, we explore the advantage of using CANDELS labeled data (KA15) to train a model and use the obtained parameters as initial condition for further training using the CL-eye labeled dataset.

\begin{figure}
  \centering
\includegraphics[width=8cm]{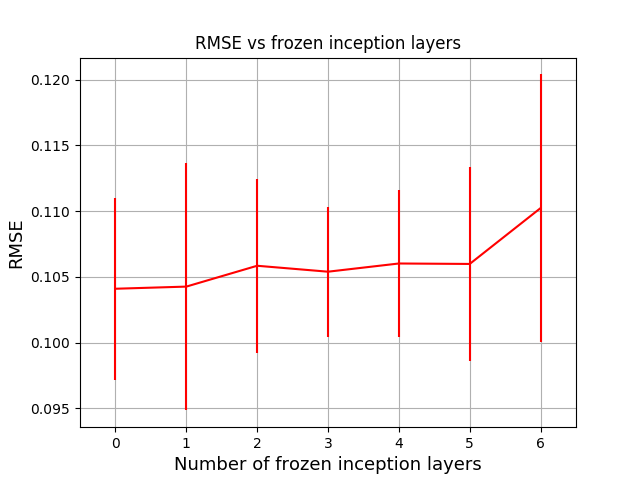}
  \caption{RMSE of our model in terms of the inception layers frozen for fine-tunning after transferring the model trained on CANDELS data. To obtain the error we freeze everything before the inception layer is indexed in the x-axis and fine-tuned over the layers after. We trained 20 models using a training set of 1,000 galaxies, and 150 galaxies for validating and testing. Error bars show standard deviation of the RMSE over these 20 cross-validation experiments.}
  \label{freez}
\end{figure}

We started by training our models using KA15 and fine-tunning some parameters using the CL-eye dataset. We directly transfered all the parameters from the model trained on KA15 (shown in Figure \ref{loss}) and we train models with CLASH over copied parameters. We evaluated freezing layers before different inception units. Figure \ref{freez} shows the effect of freezing different numbers of layers. Here, fine-tunning was performed using 1,000 galaxies in the training set, and 150 for both the validation and test sets. It can be seen that there is no statistically significant difference on the number of layers frozen. We decided to not freeze any layer, i.e. we use the model trained on KA15 as initial condition to the model trained over CL-eye, backpropagating the error through all layers. Hereafter, all models are trained without freezing any layer.

Figure \ref{lossvssize} shows the effect of the training set size used for fine-tunning over the model trained using CANDELS images with KA15 labels. As the training sample size increases, the RMSE goes down. Furthermore, the fine-tuned model outperforms the model trained over CL-eye labels with no fine-tunning. Using a training set of more than 300 galaxies does not achieve a significant improvement.

\begin{figure}
  \centering
\includegraphics[width=8cm]{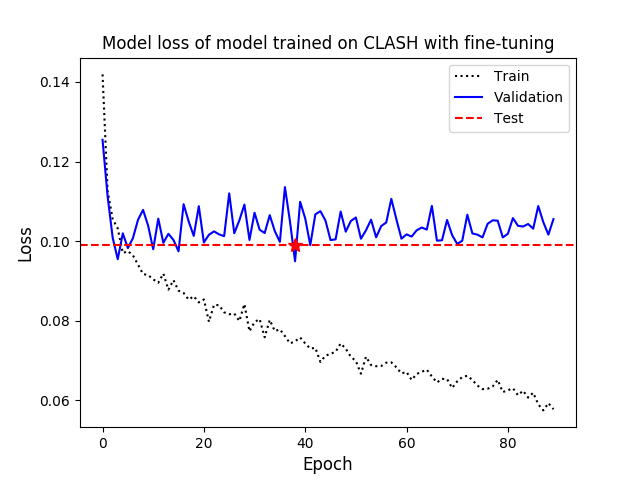}
  \caption{Learning curve of the trained CLASH using prior knowledge from CANDELS. The best model was chosen using the model in which the error on the validation set was lower. The black dotted line shows the loss over the training set, the blue solid line shows the loss over the validation set and the dashed red line shows the final test set loss over the best validation set model. A red star is shown at the epoch such model was obtained.}
  \label{loss_clash}
\end{figure}

\begin{figure*}
\begin{tabular}{ccc}
\includegraphics[width=6cm]{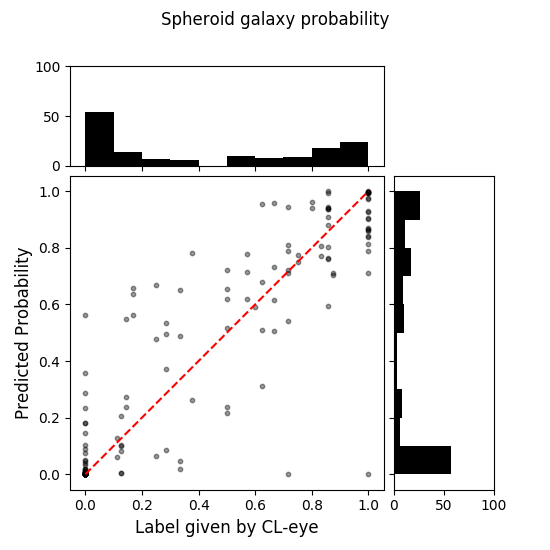} & \includegraphics[width=6cm]{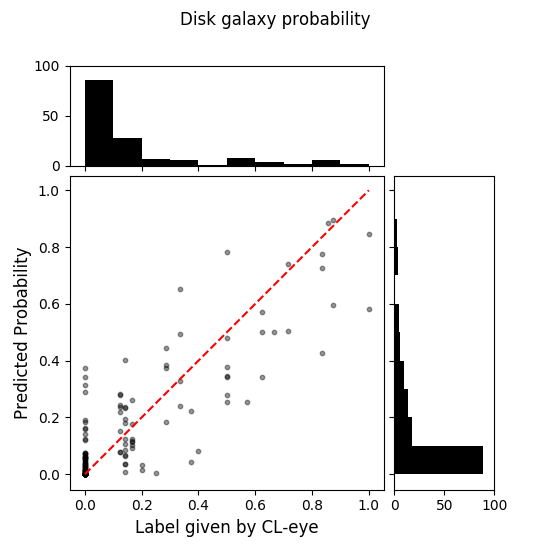} & \includegraphics[width=6cm]{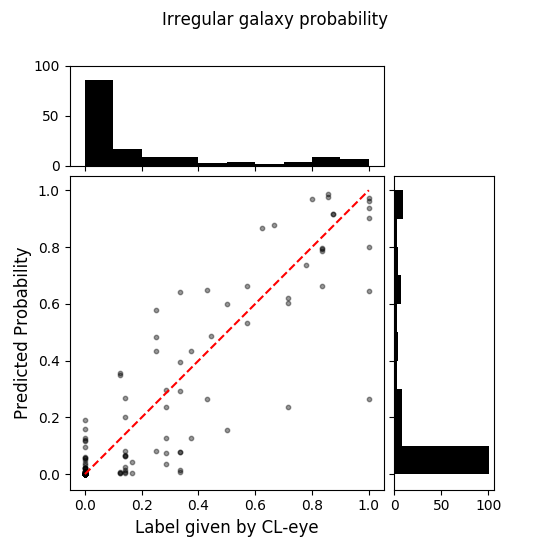}\\
\multicolumn{3}{c}{
\begin{tabular}{cc}
\includegraphics[width=6cm]{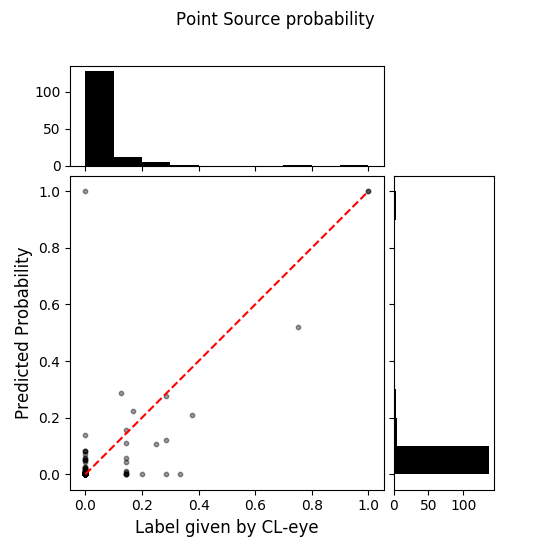} & \includegraphics[width=6cm]{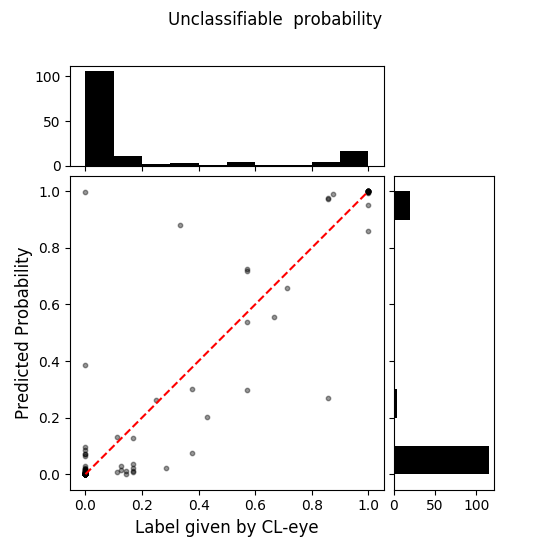}
\end{tabular}
}
\end{tabular}
\caption{Predicted probabilities from our model versus CL-eye visual labels.}
\label{pred_vs_label}
\end{figure*}

\begin{figure*}
  \centering
\includegraphics[width=18cm]{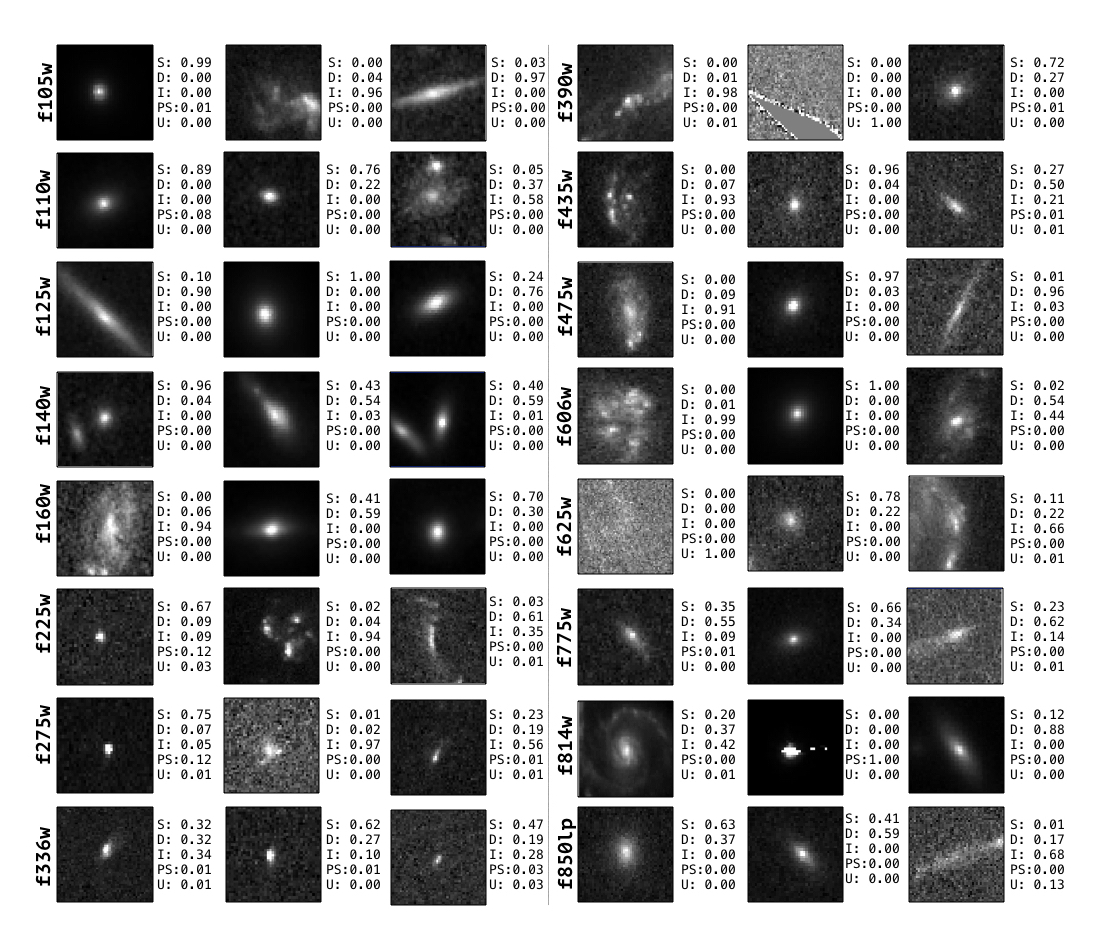}
  \caption{Images and labels of galaxies from CLASH as labeled by our model for different filters.}
  \label{labeled_images}
\end{figure*}

Figure \ref{loss_clash} shows the learning curve for the model used to produce the online catalog made publicly available with this paper. This model was the one that achieved the best RMSE over the validation dataset using a training set size of 1,000 galaxies from CL-eye fine-tuned data and without freezing. This model obtained a validation RMSE of $0.0949$ and a test RMSE of $0.0991$. In Fig \ref{pred_vs_label} we can see predicted probability values for 150 galaxies in the test set versus labels given by CL-eye.
A random subset of galaxies labeled by our model in different filters are shown in Figure \ref{labeled_images}. In this image ``S'' represents spheroid, ``D'' disk, ``I'' irregular, ``PS'' point source, and ``U'' unclassifiable.

\subsection{Using ConvNet predictions for classification}

An important use of our catalog is creating labels following a classification scheme. Here, we evaluate how our model performs using maximum a posteriori (MAP) decision rule (i.e. choosing the class with the highest probability given by our model), or choosing a threshold on the probabilities.

\begin{figure}
  \centering
\includegraphics[width=8cm]{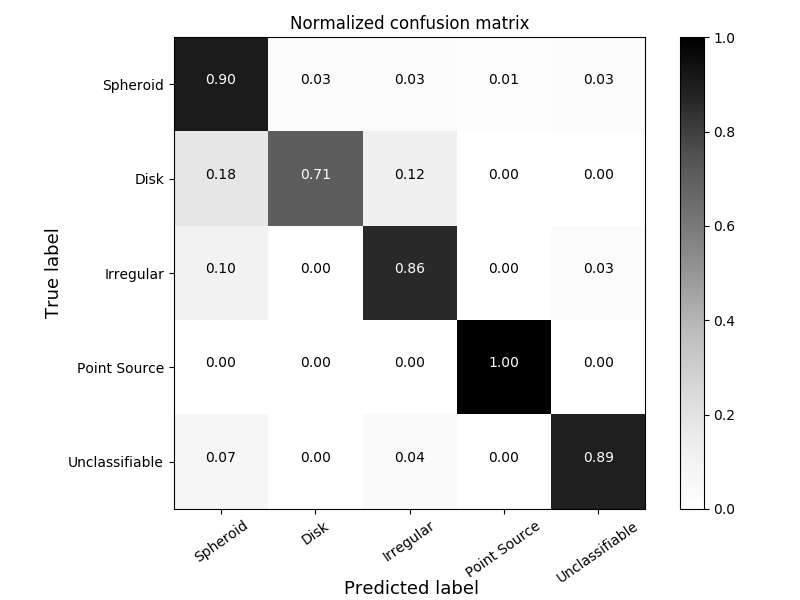}
  \caption{Confusion matrix computed on test set using the model in Fig. \ref{loss_clash}, utilizing MAP a decision rule for the probability of been spheroid, disk, irregular, point source or unclassifiable given by the model in x-axis and given by labels of CL-eye in y-axis.}
%
  \label{cmatrix}
\end{figure}

\begin{figure}
  \centering
\includegraphics[width=8cm]{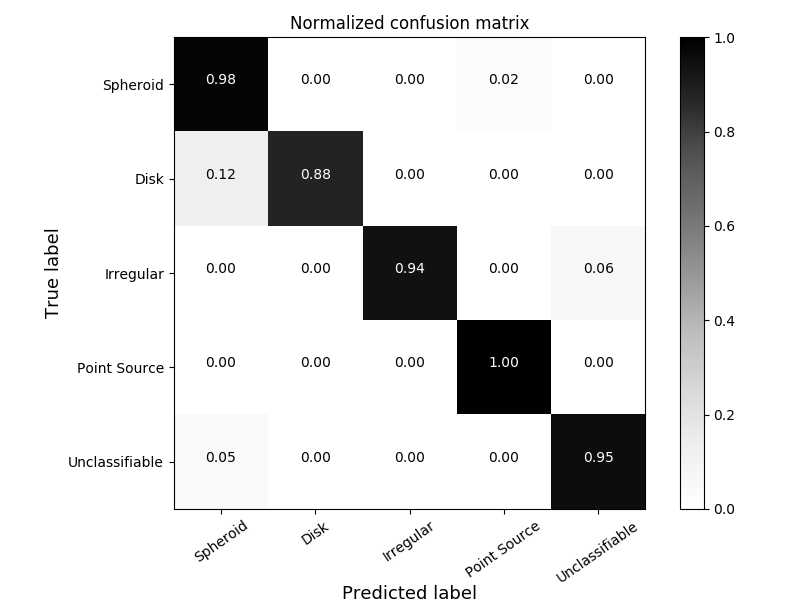}
  \caption{Confusion matrix computed on test set using the model in Fig. \ref{loss_clash}, utilizing a threshold of 0.75 on the probability of been spheroid, disk, irregular, point source or unclassifiable given by the model in x-axis and given by labels of CL-eye in y-axis.}
%
  \label{cmatrix_075}
\end{figure}

\begin{figure}
  \centering
\includegraphics[width=8cm]{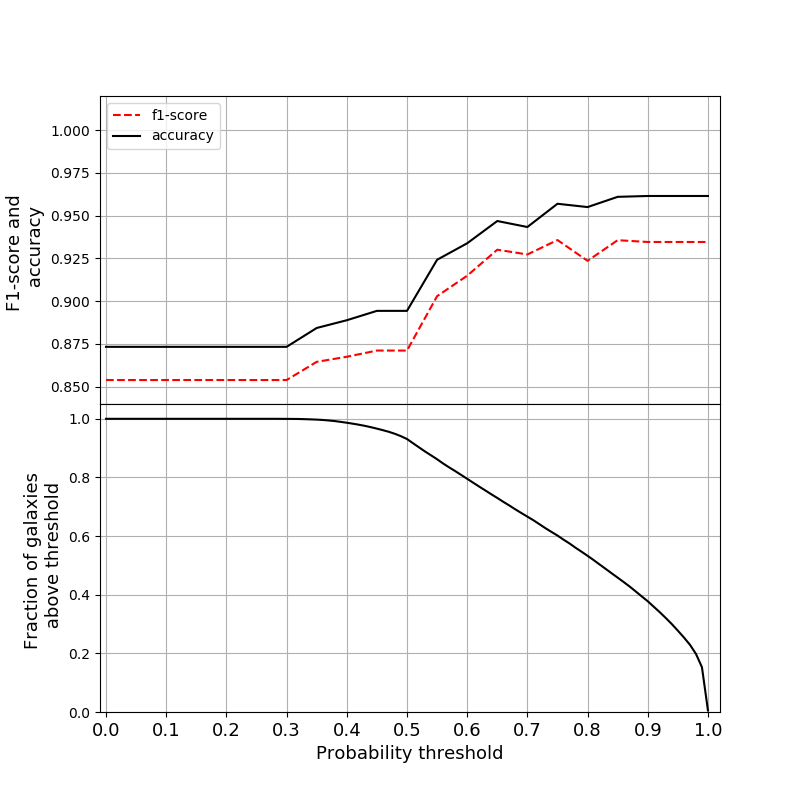}
  \caption{Accuracy, macro f1-score, and fraction of galaxies in terms of the probability threshold over our model's predictions used to define a class. As we increase the probability threshold, the accuracy and f1 score increases, and the number of galaxies with a single probability over this threshold diminishes.}
  \label{fig:n_vs_pbb}
\end{figure}

Figure \ref{cmatrix} shows the confusion matrix of our model following a MAP decision rule classification scheme. Here, we assign to each object the class that has the highest probability. The worst performance occurs when our model confuses $\sim 24\%$ of the disk galaxies with spheroids. By using this scheme, we correctly label 91\% of spheroids, 65\% of disks, 73\% of irregulars, 100\% of point sources, and a 89\% of unclassifiable sources. \cite{marc} propose to select a threshold of 0.75 on the probabilities in order to have a higher certainty on the labels. This is something important to consider as human labels are not perfect. We explore this idea by evaluating our model against galaxies with high certainty on their human labels. Figure \ref{cmatrix_075} shows the confusion matrix using a threshold of 0.75 in the human labels probabilities to define the classes. In this experiment our model correctly labels 98\% of spheroids, 100\% of disks, 94\% of irregulars, 100\% of point sources, and 100\% of the unclassifiable sources. We further explore these results by assessing the performance of our model in terms of the probability threshold used to classify sources. Figure \ref{fig:n_vs_pbb} shows the accuracy and macro f1-score in terms of the probability threshold used to decide the class of each object for our test set. Galaxies with a probability lower than such threshold for all classes are not considered. As expected, as the threshold grows, the performance of our model tends to be better. There is a collateral effect to choosing such threshold: as this value increases, we are considering less galaxies. Figure \ref{fig:n_vs_pbb} shows the impact of  increasing the probability threshold on the number of selected galaxies. We keep around a 80\% of the galaxies for a threshold of 0.60 on the probability, corresponding to an f1-score of approximately 0.91 and an accuracy of 0.94. If we use a threshold on the probabilities of 0.90, we keep approximately a 40\% of the galaxies, but the f1-score and accuracy are higher than 0.93.







\section{CATALOG}

This paper is accompanied by a visual-like  catalog containing morphology probability predictions for $68,531$ images of $8,532$ galaxies in 16 HST photometric bands for galaxies in the 25 CLASH fields. Morphologies correspond to the probabilities predicted by our ConvNet model.
We give the probability for each galaxy of being disk, spheroid, point source/compact, peculiar/irregular or unclassifiable in each of the 16 different HST photometric bands (F225W, F275W, F336W, F390W, F435W, F475W, F606W, F625W, F775W, F814W, F850LP, F105W, F110W, F125W, F140W, F160W), with their corresponding labels, cluster name, right ascension, and declination for each image where the galaxy appears. 
We provide a sample table in Table \ref{catalog}, while the full table in machine-readable format is available in \url{http://www.inf.udec.cl/~guille/data/Deep-CLASH.csv}. The catalog provides the following information:

\begin{description}
\item[Cluster:] Cluster name to which a galaxy belongs;
\item[ID:] ID of the galaxy as reported in Molino et al.'s catalog.
\item[RA:] Right ascension of the galaxy (J2000);
\item[DEC:] Declination of the galaxy (J2000);
\item[F225w\_PSP:] probability for an object to be a spheroidal galaxy in the indicated band (from F225W to F160W);
\item[F225w\_PD:] probability for an object to be a disk galaxy in the indicated band (from F225W to F160W);
\item[F225w\_PI:] probability for an object to be an irregular galaxy in the indicated band (from F225W to F160W);
\item[F225w\_PPS:] probability for an object to be a point source in the indicated band (from F225W to F160W);
\item[F225w\_PUN:] probability for an object to be unclassifiable in the indicated band (from F225W to F160W);
\item[...]
\end{description}

\begin{table*}
\begin{center}
    \begin{tabular}{ | c | c | c | c | c | c | c | c | c | c |}
    \hline
    Cluster & ID & RA & DEC & F225w\_PSP & F225w\_PD &F225w\_PI &... & F160w\_PPS & F160w\_PUN \\ \hline
     ms2137&$287$& $325.0698$ &$-23.6508$ &$0.00$ & $0.01$&$0.77$ &... &$0.99$ & $0.00$ \\ 
     macs1931&$1708$&$292.9468$&$-26.5854$&$0.13$&$0.12$ &$0.28$&.. &$0.99$ &$0.00$  \\ 
     rxj1347&$16$&$206.8822$ & $-11.7663$  &-  & - & - &...  & $0.00$ & $0.00$   \\
     ... & ...  & ... &...  & ...  & ... &...  &...&...&...\\
     rxj2248 &$43$& $342.1780$ & $-44.5120$  &$0.00$ & $0.00$ &$0.00$ &...  & $0.00$ & $0.06$  \\ \hline
    \end{tabular}
  \caption{Visual-like morphology catalog of galaxies over CLASH produced by the ConvNet trained in CLASH sample with fine tuning. Full version available in \url{http://www.inf.udec.cl/~guille/data/Deep-CLASH.csv}. Table made to guide to the readers.}
  \label{catalog}
\end{center}
\end{table*}

\section{SUMMARY AND CONCLUSIONS}
This paper presents a visual-like morphological classification of $8,532$ galaxies in 16 HST photometric bands of CLASH, one morphology per filter, for a total of $68,531$ labeled images. The catalog contains probabilities of being spheroid, disk, irregular, point source or unclassifiable, predicted using a convolutional neural network (ConvNet) model. The model \footnote{\textbf{A demo on how to use our model to predict galaxy morphologies in CLASH is publicly available at \url{https://github.com/manacho14/CLASH.git.}}} was trained using $\sim 7,500$ CANDELS images and further fine-tuned using CLASH images with eyeball labels. We show that by using more than 300 CLASH labeled images to train the model, it does not improve significantly in terms of cross-validated root-mean-square-error. We evaluate how the model performs on classification tasks in terms of the probability threshold used to define the class. The higher the threshold, the better performance of the model in terms of f1-score and accuracy, but the less galaxies we keep. In that sense, there is a trade-off between the certainty on the labels and the number of galaxies that can be used, which has to be taken into account when using the catalog. We believe that the fine-tuning techniques described in this work are essential to label data from new instruments such as the Large Synoptic Survey Telescope and EUCLID when new data are available with a limited amount of labeled data.

This catalogue constitutes a complement to the automated morphology in CANDELS since it provides morphologies for galaxies in clusters. The comparison between cluster and field galaxies is fundamental to study the effects of the environment in the morphological transformation of galaxies (e.g. formation of early-type galaxies) and its relationships with the evolution of star formation and stellar mas build-up. Furthermore, the catalogue is the first multi-wavelength morphological catalogue, which can be employed in the search of morphologically peculiar galaxies, which are likely undergoing interactions with their surrounding environment. It is therefore an important resource for the study of galaxy interactions.

\section{Acknowledgment}
Support for this work was provided by NASA through grant number from the Space Telescope Science Institute, which is operated by AURA, Inc., under NASA contract NAS 5-26555. This work is based on observations taken by the CANDELS Multi-Cycle Treasury Program with the NASA/ESA HST, which is operated by the Association of Universities for Research in Astronomy, Inc., under NASA contract NAS5-26555. This publication uses data generated via the Zooniverse.org platform, development of which is funded by generous support, including a Global Impact Award from Google, and by a grant from the Alfred P. Sloan Foundation. G.C.V. acknowledges the support provided by FONDECYT postdoctoral research grant no 3160747; CONICYT-Chile and NSF through the Programme of International Cooperation project DPI201400090; the Ministry of Economy, Development, and Tourism's Millennium Science Initiative through grant IC120009, awarded to The Millennium Institute of Astrophysics (MAS). P.C. acknowledges the support provided by FONDECYT postdoctoral research grant no 3160375. R.D. gratefully acknowledges the support provided by the BASAL Center for Astrophysics and Associated Technologies (CATA). We acknowledge Crist\'obal Donoso-Oliva, Hugo Parischewsky-Zapata and Daniela Olave-Rojas who helped label CLASH images.




\bibliographystyle{mnras}
\bibliography{refs} 



\bsp	
\label{lastpage}

\end{document}